\documentclass[traditabstract]{aa} 
\usepackage{graphicx}
\usepackage{longtable}

\begin{document}
   \title{Decimetre dust aggregates in protoplanetary discs}


   \author{J. Teiser
          \inst{1}
          \and
          G. Wurm
          }

   \institute{Institut f\"ur Planetologie, Wilhelm-Klemm-Str. 10, D-48149 M\"unster, Germany\\
              \email{j.teiser@uni-muenster.de}}

   \date{Received March 11, 2009}

 
  \abstract
   {The growth of planetesimals is an essential step in planet formation. Decimetre-size dust agglomerates mark a transition point in this growth process.
   In laboratory experiments we simulated the formation, evolution, and properties of decimetre-scale dusty bodies in protoplanetary discs.
Small sub-mm size dust aggregates consisting of micron-size SiO$_2$ particles randomly interacted with dust targets of varying initial conditions in a continuous sequence of independent collisions. Impact velocities were 7.7 m/s on average and in the range expected for collisions with decimetre bodies in protoplanetary discs.
   The targets all evolved by forming dust \emph{crusts} with up to several cm thickness and a unique filling factor of \mbox{31 \% $\pm$3 \%.} A part of the projectiles sticks directly. In addition, some projectile fragments slowly return to the target by gravity. All initially porous parts of the surface, i.e. built from the slowly returning fragments, are compacted and firmly attached to the underlying dust layers by the subsequent impacts. Growth is possible at impact angles from 0$^{\circ}$ (central collision) to 70$^{\circ}$. No growth occurs at steeper dust surfaces. We measured the velocity, angle, and size distribution of collision fragments. The average restitution coefficient is 3.8 \% or 0.29 m/s ejection velocity. Ejecta sizes are comparable to the projectile sizes.
   The high filling factor is close to the most compact configuration of dust aggregates by local compression ($\sim 33$\%). This implies that the history of the surface formation and target growth is completely erased. In view of this, the filling factor of 31\% seems to be a universal value in the collision experiments of all self-consistently evolving targets at the given impact velocities. We suggest that decimetre and probably larger bodies can simply be characterised by one single filling factor. While gravity dominates re-accretion in the experiments, small fragments will be re-accreted as well in protoplanetary discs by gas drag at the given low ejection velocities. The accretion efficiency in planetesimal growth is model dependent. However, a small fraction of small particles re-accreted by gas flow or direct sticking readily allows growth of dusty bodies in protoplanetary discs in the decimetre range.}


   \maketitle
%

\section{Introduction}

   The formation of planetary bodies still has a number of unsolved problems, such as how kilometre-size bodies (planetesimals) form.  While it is widely accepted that planetesimals grow within a protoplanetary disc consisting of gas and dust, it remains an open question which processes lead to their formation. There are two main formation scenarios.

One scenario for planetesimal formation is based on gravitational attraction of solid bodies within the protoplanetary disc including gravitational instabilities within dense regions. The idea of a dust laden sub layer within the disc leading to a gravitational collapse was first published by Safronov (1969) and Goldreich \& Ward (1973). Later it was shown by e.g. Weidenschilling et al. (1989) or by Schr\"apler \& Henning (2004) that shear induced turbulence is sufficient to prevent the formation of a dense enough sub-layer. Under certain circumstances turbulence might create high densities of solids by concentrating large boulders within eddies which then can get gravitational instable (Johansen et al. 2006). This model requires a significant mass to be in decimetre-size bodies as an initial reservoir. The properties of decimetre bodies are therefore of importance for further evolution. Property here essentially means porosity or internal morphology, which determines the outcome of further encounters between aggregates.

Decimetre bodies are assumed to be formed by coagulation of smaller aggregates. Coagulation is the second process considered for the formation of planetesimals. This scenario is based on growth by mutual collisions and sticking of dusty bodies all the way up to km-size objects. Decimetre-size aggregates are of special importance in this scenario as well. At decimetre-size collision velocities increase strongly up to 60 m/s (Weidenschilling and Cuzzi 1993, Sekiya and Takeda 2003). If further growth is possible depends on the properties of the decimetre-size dust aggregates. Wurm et al. (2005a) found that highly porous (granular) dust aggregates are destroyed by impacts of sub-cm bodies at large impact velocities. In contrast, Teiser and Wurm (2009) and Wurm et al. (2005b) showed that growth up to $\sim$ 60m/s is possible if the target is compact (33\% filling factor). Therefore, for both scenarios it is important to know how decimetre bodies grow. This could not be answered so far as decimetre bodies have a complex collisional history and the morphology might evolve complex as well. This is the focus of this paper. 

Coagulation models for aggregate growth in protoplanetary discs are based on the interactions of solids with the gas as a motor for relative velocities between particles. Several authors have modeled coagulation processes within protoplanetary discs by different methods (e.g. Weidenschilling \& Cuzzi 1993, Weidenschilling 1997, Ormel et al. 2008, Brauer et al. 2008).

The velocities of solid bodies, which lead to collisions, depend on the disc properties, as for example the gas pressure profile, the gas to dust ratio or the strength of turbulence. Further, particle properties such as density, porosity or size determine the relative velocities between different particles. In the different models more or less plausible assumptions are made for the results of particle collisions, for example perfect sticking or completely destructive collisions with power law distributions for the fragments. As these assumptions are a central part of the models and cover a wide range of different collision dynamics the resulting predictions for the possibility of planetesimal growth differ in the same wide range. 

As the outcome of collisions within protoplanetary discs has a strong influence on the results of coagulation models, studies on the physics of collisions are crucial to get plausible conclusions from these models. Within the last years it was shown experimentally and theoretically that millimetre to centimetre-size aggregates form on very short timescales independent from the disc model (Dominik \& Tielens 1997, Blum et al. 2006, Blum \& Wurm 2008). Icy and/or organic materials might aid these growth processes by enhancing the stickiness of aggregates (Bridges et al. 1996, Kouchi et al. 2002). With increasing aggregate size, velocities between different size aggregates become larger, bodies get more and more compact and bouncing and fragmentation play an important role in further collisions. 

Depending on the disc model, collision velocities for m-size bodies reach values of several 10 m/s. In the past this often has been treated as an obstacle to planetesimal growth as collisional sticking seems to be unlikely in this parameter range. However, growth by collisions at large velocities is possible as Wurm et al. (2005b) and Teiser \& Wurm (2009) showed for velocities of up to 56.5 m/s in experiments with highly compressed dust agglomerates. They suggest that fragmentation is the key process for growth in this regime as it dissipates efficiently a large part of the impact energy. 

In addition, different mechanisms have been proposed to enable or enhance growth in the fragmentation dominated regime. One effect is collisional charging together with electrostatic re-accretion of fragments (Poppe et al. 2000, Blum 2004). It has been proposed that during an impact the ejected fragments are charged complementary to the target surface which then leads to an attractive force and enables the re-accretion of the charged fragments. 

Another important mechanism to aid growth is re-accretion of ejecta by gas flow as proposed by Wurm et al. (2001a,b) and also analysed by Sekiya and Takeda (2003). Projectiles hitting the target surface bounce off or fragment and small particles couple to the surrounding gas rapidly. If they do so fast enough the gas flow will drag them back to the surface. This process is limited to objects smaller than several metres where the gas flow is not yet in the hydrodynamic limit. Otherwise particles are transported around the body rather than back to the surface. However, especially for cm-dm bodies this effect can efficiently re-accrete ejecta, which eventually can stick at secondary, much slower impacts. We will refer to this process later on as its effects and strength are similar to the role of gravity in our experiments. 

The outcome of collisions is determined by the mechanical properties of the colliding particles which are mainly dominated by the porosity. A number of experimental studies was carried out with velocities far below 10 m/s using highly porous dust agglomerates with a volume filling (volume fraction of a dust agglomerate filled with material) of only $\sim$10\% (Blum et al. 2006, Blum et al. 2004). At low speed below 1 m/s, mm-size particles of similar size only rebound in mutual collisions. (Hei{\ss}elmann et al., unpublished).  The mm-size projectiles can add their mass though to a larger target in collisions up to 3 m/s (Langkowski et al. 2008). Sticking collisions lead to a compaction of the targets as the projectile buries itself within. Compaction by low speed collisions has also been observed by Weidling et al. (2009). 

For more compact aggregates no sticking has been found in collisions of similar size mm particles up to 4 m/s (Blum \& M\"unch 1993). It is unclear yet how particles grow from a few mm to a few cm in a velocity range up to a few m/s. However, we note that the experiments reported here also show growth at collision velocities below 1m/s in some aspects. This is not a focus here but supports the idea that a size of a few cm can be reached by collisions somehow. We therefore assume here that a fraction of the particles has grown to a few cm. 

Velocities around 10 m/s are expected for decimetre-size bodies. In a cometesimal formation (coagulation) model by Weidenschilling (1997) decimetre bodies mostly gain their mass by impacts with aggregates of tens of microns in size. In experiments with larger velocities larger than \mbox{9 m/s} and using compact dust aggregates with a volume filling of \mbox{$\sim$33\%}, sticking is observed together with fragmentation (Wurm et al. 2005b, Teiser \& Wurm 2009). In similar collision experiments at comparable velocities but with a porous target only fragmentation was observed (Wurm et al. 2005a). Due to the fragmentation, sticking and bouncing it is not clear how such a body evolves further. 

To shed some light on this size-scale we carried out experiments in which we determined the properties of dust agglomerates which formed by collisional accretion and re-accretion of sub-mm size particles with impact velocities of 7.7 m/s on average. Former studies mostly dealt with collisions as single events. There are always some fluctuations in the mechanical properties of dust agglomerates so there is always uncertainty in scaling from individual collisions to the net effect of a statistically large number of impacts. Within this work we do not discuss impacts as single events but gain our information by analysing the net outcome of a large number of individual collisions. Fluctuations in the mechanical properties of the agglomerates are levelled out by statistics in literally millions of collisions. The experiments presented here allow conclusions about formation and evolution of decimetre dust agglomerates in protoplanetary discs.


\section{Experiment description}

\begin{figure}
\includegraphics[width=9cm]{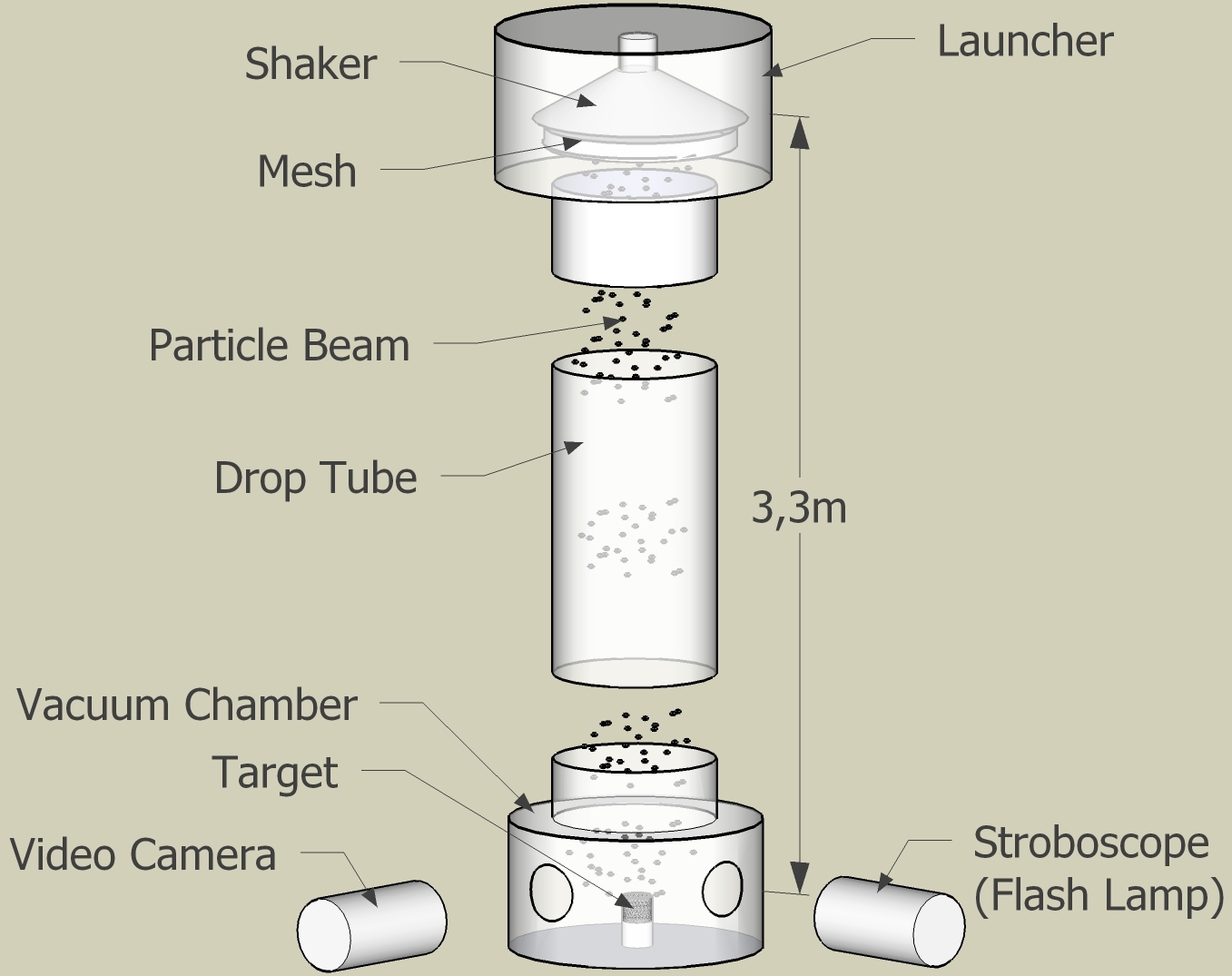}
\caption{The experiment consists of a dust generator, the drop tube acceleration, and the impact chamber. Different targets can be placed inside the impact chamber. The collisions are visualised by a video camera and stroboscopic illumination at 500 Hz with a xenon flash lamp.}
\label{aufbau}
\end{figure}

%
We developed an experiment in which sub-mm dust aggregates can be launched continuously and which allows study of the growth of a massive dusty body. In fig. \ref{aufbau}. the basic components of our experimental setup are shown schematically. The setup can be divided in three parts: 
the dust generator, the accelerator, and the impact chamber.

\textbf{Dust generator:} To launch particles continuously in large amounts, we generate a beam of dust aggregates by vibrating a dust powder sample placed on a sieve with given mesh-size. The frequency, amplitude, and duty cycle of the dust generator can be adjusted and were set to values where a dense dust beam forms. For the experiments reported here, we use a mesh with 250 $\mu$m holes. 

As dust we used a commercial quartz powder as used in our earlier studies (Wurm et al. 2005b, Teiser \& Wurm 2009). The typical size of dust grains is between 1$\mu$m and 5$\mu$m. The size-distribution of projectiles (dust aggregates) has been measured by microscopy of particles on a surface placed directly below the sieve under ambient pressure. As launched particles drop in air and only for a short distance they sediment to the surface slowly at sub-m/s. Particles do not fragment or change their size significantly at these speeds (Wurm et al. 2005b). We also measured the size-distribution of particles after being accelerated in the drop tube and after impact at several m/s. Here a target of compressed graphite powder was placed in the vacuum chamber for a colour contrast between target and projectile material. Compressed graphite has a similar porosity as compressed quartz dust. The mechanical properties of dust aggregates are mainly  determined by the porosity of the dust agglomerate as shown by Langkowski et al. (2008) and Blum et al. (2006), so we assume the mechanical properties to be comparable. The results of these impacts also were analysed by microscopy. 

From the microscope images we define the size of a particle as radius of a sphere with equivalent cross section. Fig. 2 gives the normalised mass (or volume (size$^3$)) distribution of the impacting projectiles and the particles after impact. We note that particles can be larger than 125 $\mu$m in effective radius though the sieve mesh has 250 $\mu$m opening. A sieve only restricts particles in two dimensions but still allows elongated particles to have a larger third dimension. 

In agreement to earlier studies at similar velocities the sizes of the particles do not change significantly due to the impacts (Wurm et al. 2005b). We see a slight fragmentation on impact as described below but this is not prominent in the size-distributions.

\begin{figure}
\begin{center}
\includegraphics[width=8cm]{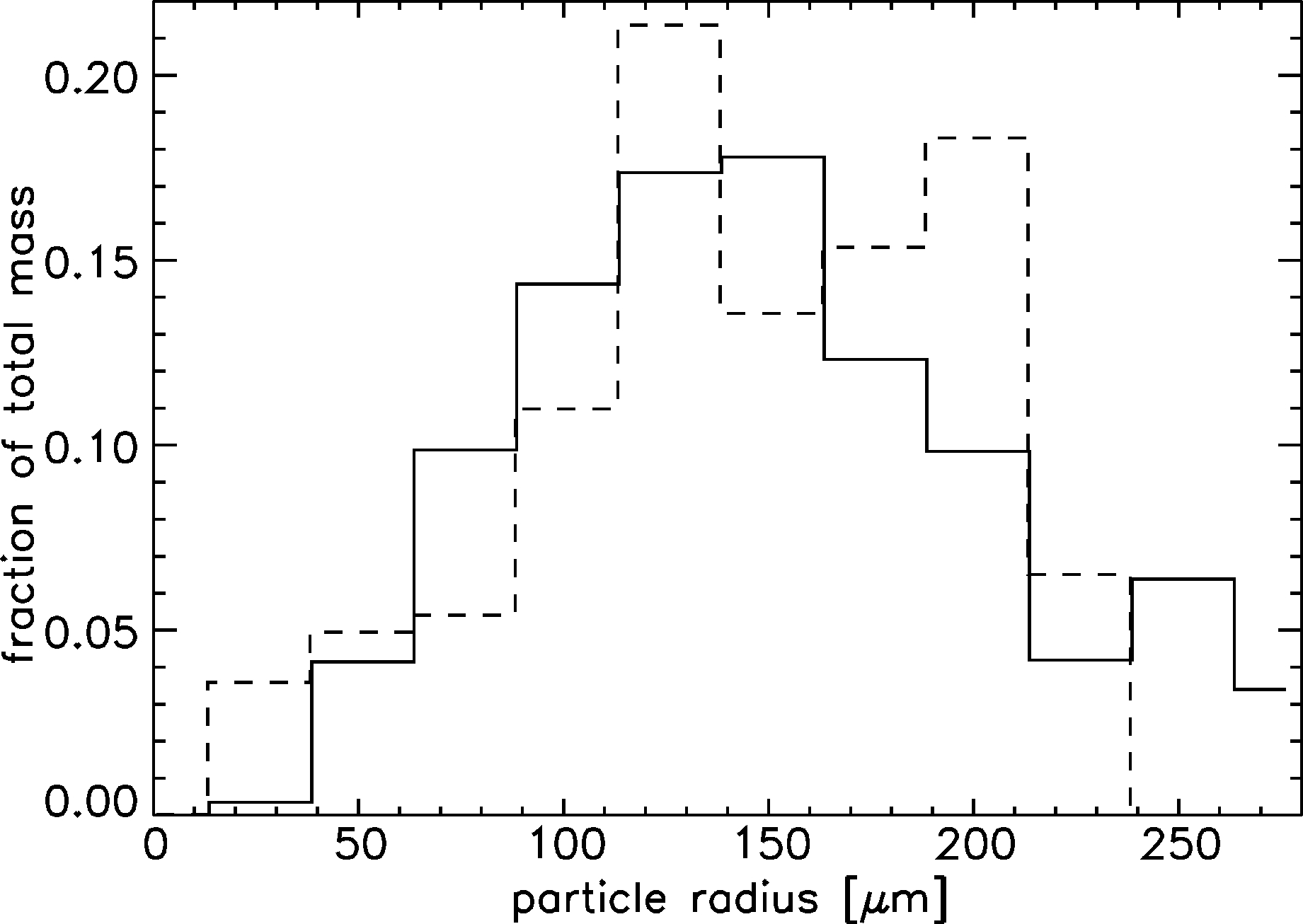}
\end{center}
\caption{Mass-distribution of particles (fraction of total mass versus particle size); solid line: impacting particles; dashed line: particles after impact.}
\label{sizes}
\end{figure}

\begin{figure}
\begin{center}
\includegraphics[width=8cm]{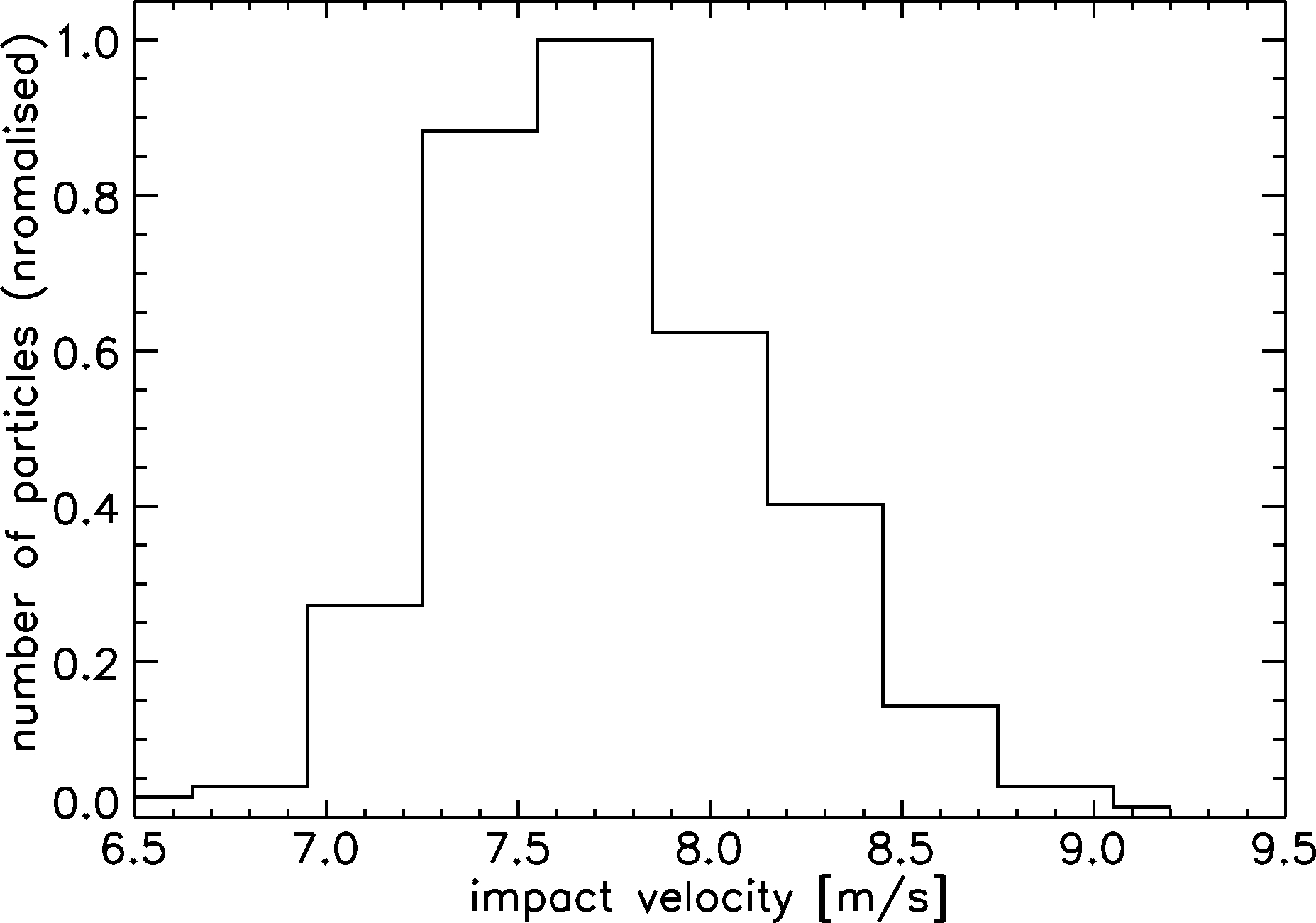}
\end{center}
\caption{Velocity-distribution of projectiles measured in the center of the impact chamber normalised to 1 at maximum.}
\label{speed}
\end{figure}

\textbf{Acceleration:} The acceleration of the particles is achieved by gravity as the dust aggregates pass a drop tube. Due to free fall particles do not change between launch and impact. The drop tube provides a free fall height of \mbox{3.3 m} and has a diameter of 15 cm. As the dust generator is as large as the drop tube, the system generates a wide beam and large targets up to decimetre-size can be impacted repeatedly in a random sequence all over the surface. Different impact angles are realised by the target surface. For the experiments the drop tube as well as the generator and the impact chamber are evacuated to below 10$^{-2}$ mbar. Gas-grain coupling times for 100$\mu$m particles are of the order of several s then. This is larger than the free fall time of 1s and particles can be considered to accelerate in free fall. A particle at rest at top of the generator would reach a final velocity at the bottom of the drop tube of about \mbox{8.0 m/s.} This is consistent with the velocity distribution measured with an average of 7.7 m/s (fig. \ref{speed}), considering that the vibrations for cloud formation result in a small non-zero initial velocity.

\textbf{Impact chamber:} The drop tube is attached to the impact chamber where different targets can be placed. The collisions can be observed by a video camera and stroboscopic illumination, which allows a velocity determination as e.g. given for impacting particles in fig. 3. Fig. 4a shows an example of imaged rebounding ejecta. As the projectile-size is of the order of the spatial resolution of our camera, we did not attempt to constrain the size of the airborne particles from these observations. Trajectories can very well be fitted by parabolas with gravitational acceleration \mbox{g=9.81 m/s$^2$} (fig. 4b). This allows the determination of the rebound speed in two dimensions and rules out other forces to be significant. E.g., Poppe et al. (2000) observed electrical charging in dust-target collisions. Blum (2004) suggested that charge build-up might lead to electrostatic re-accretion of ejected particles. As particle trajectories in our experiments perfectly fit gravitational parabolas electrostatic fields and grain charging are not important for particle motion here.

\begin{figure}
\begin{center}
\includegraphics[width=6cm]{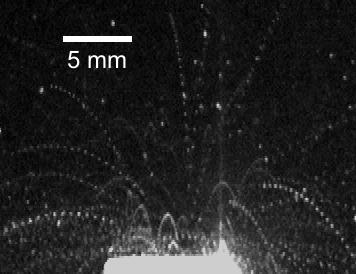}
\vspace{0.2cm}\\
\includegraphics[width=6cm]{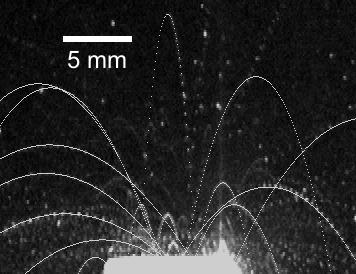}
\end{center}
\caption{Incoming projectiles and rebounding ejecta in stroboscopic illumination (a). Examples of parabolas induced by gravity which have been fitted to ejecta trajectories to determine the 2d rebound speeds (b).}
\label{impact}
\end{figure}

\section{Experimental results}

As seen in fig. 4 ejecta trajectories have been measured for aggregates impacting a dusty surface generated by the preceding impacts. Fig. 5 shows the velocity-distribution of the ejected and rebounding particles in vertical and in one (observed) horizontal direction. 

\begin{figure}
\begin{center}
\includegraphics[width=8cm]{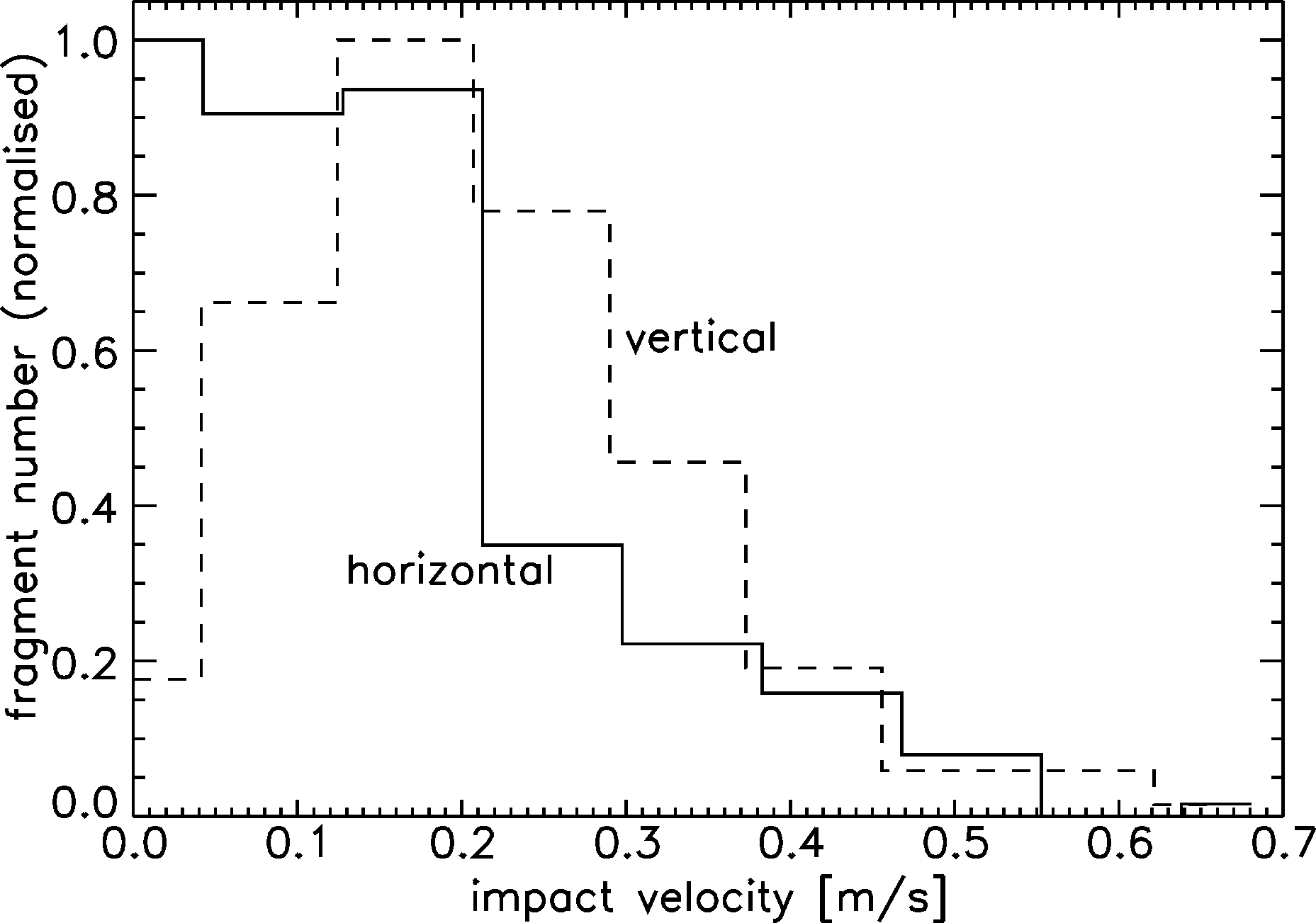}
\end{center}
\caption{Velocity distribution of ejecta of projectiles impacting a dusty surface which had been generated by previous generations of dust impacts; dashed line: vertical component; solid line: 1-d horizontal component.}
\label{fragspeed}
\end{figure}

The average velocity in vertical direction is \mbox{v$_z$ = 0.21 m/s.} The average velocity in horizontal direction is \mbox{v$_x$ = 0.14 m/s.} If both velocity components are independent the absolute value of the rebound velocity on average is \mbox{v$_{ej}$ = 0.29 m/s.}  

All targets which were placed within the chamber grew thick crusts of dust. Essentially the top surface of each target constrains the size and shape of the growing dust aggregate. The original target material is not important. After the first layer of dust the impacts can be considered as self consistent evolution of a dust target of the same dust powder as the projectiles. Table 1 gives a summary of all targets used and the properties of the resulting dust crust. 

\begin{table*}[tb]
\center
\begin{tabular}{|p{1.3cm}|p{1.5cm}|p{0.8cm}|p{0.8cm}|p{1.2cm}|p{1.4cm}|p{1.8cm}|}\hline
target base & size & dust mass (g) & dust volume (cm$^3$) & volume filling (\% ) & maximum impact angle for growth & miscellaneous\\  \hline
cylinder & 9.50 mm diameter & 0.166 & 0.197 & 32.4 $\pm$ 1 & 65.3 $\pm$ 1
$^{\circ}$ & circular cross section for growth\\ \hline
cylinder & 6.40 mm diameter & 0.055 & 0.0704 & 30.0 $\pm$ 1 & 65.3 $\pm$ 1
$^{\circ}$ &  \\ \hline
glass plate & 1.1 mm $\times$ 28 mm & 0.016 & 0.0233 & 26.4 $\pm$ 1 & 70.3 $\pm$ 1$^{\circ}$ & \\ \hline
cylinder (horizontal) & 13 mm $\times$ 28 mm mantle area & 1.190 & 1.46 & 31.7 $\pm$ 4 & - & rotating cylinder, growth on mantle\\ \hline
glass beaker & 50 mm (85 mm height) & 40.037 & 52 & 29.6 $\pm$ 1 & - & growth within beaker\\ \hline
cylindrical aluminum tray & 30.1 mm (10 mm height) & 10.818 & 12.05 & 34.5 $\pm$ 2 & 68 $\pm$ 1$^{\circ}$ & growth beyond tray height\\
\hline
\end{tabular}
\caption{Properties of dust aggregates grown on specific targets}\label{tab}
\end{table*}

Examples of grown dust targets can be seen in fig. 6. Targets grow as long as the envelope is not steeper than 70$^{\circ}$ (fig. 6, top left). The other targets in fig. 6 are not yet fully grown end configurations as the angle still decreases from the center to the edge. To determine the filling factor we measured the mass of the grown dust and determined the volume. For Fig. 6 we fitted parabolic profiles which fit the data well and assume rotational symmetry to determine the volume. For fig. 6 (top left) we assume a triangular profile.

\begin{figure}
\begin{center}
\includegraphics[width=8cm]{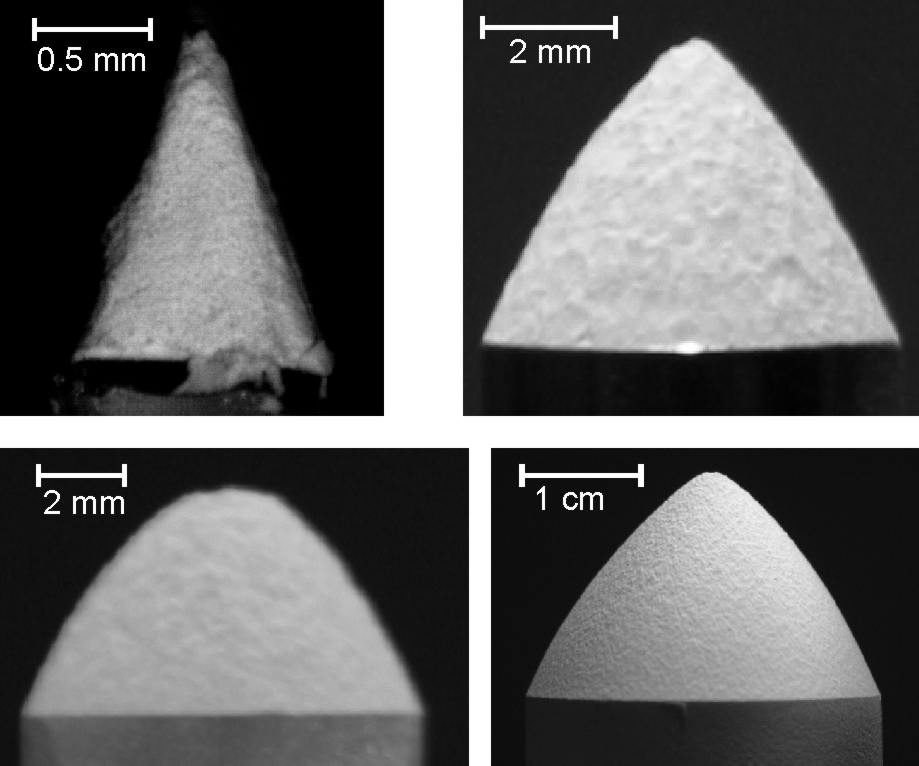}
\end{center}
\caption{Dust targets grown by the impacts; top left: a thin (1.1 mm) edge of a glass plate; other targets: cylinders of different diameter.}
\label{kegel}
\end{figure}

The dust added to a target can be placed in two categories.
\begin{itemize}
\item Projectile parts sticking directly during a primary impact.
\item Ejecta returning by gravity and sticking in secondary, slower collisions.
\end{itemize}

Direct sticking of dust aggregate parts was the center of earlier experiments and e.g. observed by Wurm et al. (2005b) at velocities down to 9 m/s. Within the resolution of our optical system we also see particles sticking at impact sites here at about 7.7 m/s. At much smaller impact velocities corresponding to our ejecta velocities of the order of 1 m/s or below, dust sticking also occurs in the experiments reported here. This is immediately visualised as we see dust sticking on vertical parts of our targets if these are protected from fast eroding vertical impacts (see below). In video microscopy of a target surface (not shown here) we also saw particles of a growing dust surface being removed by impacts. 

The second class of particles added to a target are slow ejecta at velocities of 1 m/s and less (see fig. 4 and fig. 5.) which return to a target by gravity. The role of gravity in the re-accretion and the relation to dust growth in protoplanetary discs will be discussed further below. Which fraction of ejecta is re-accreted depends on the size of the target. For small targets only slow ejecta can return while fast ejecta miss the target (see fig. 4). The fraction of re-accreted ejecta is clearly visible in our experiments as the dust growth in the center of a target is enhanced (see fig. 6). The incoming projectile beam can be regarded as homogeneous over the width of the targets and an inhomogeneous growth especially one which is symmetric on all targets has to be due to re-accreted ejecta. Enhanced growth in the center of a target implies that, statistically, it is more likely that ejected particles reach the center of a target than the edge. At the measured ejection velocity in z-direction the typical particle can rise again to a height of 3 mm. Even fast (1 m/s) ejecta can only rise about 5 cm (see fig. 4). Our targets were placed at larger heights of more than 10 cm (except for the chamber bottom). Particle growth on the target is therefore only influenced by impacts hitting the target, not ejecta leaving the bottom of the chamber. 

\begin{figure}
\begin{center}
\includegraphics[width=9cm]{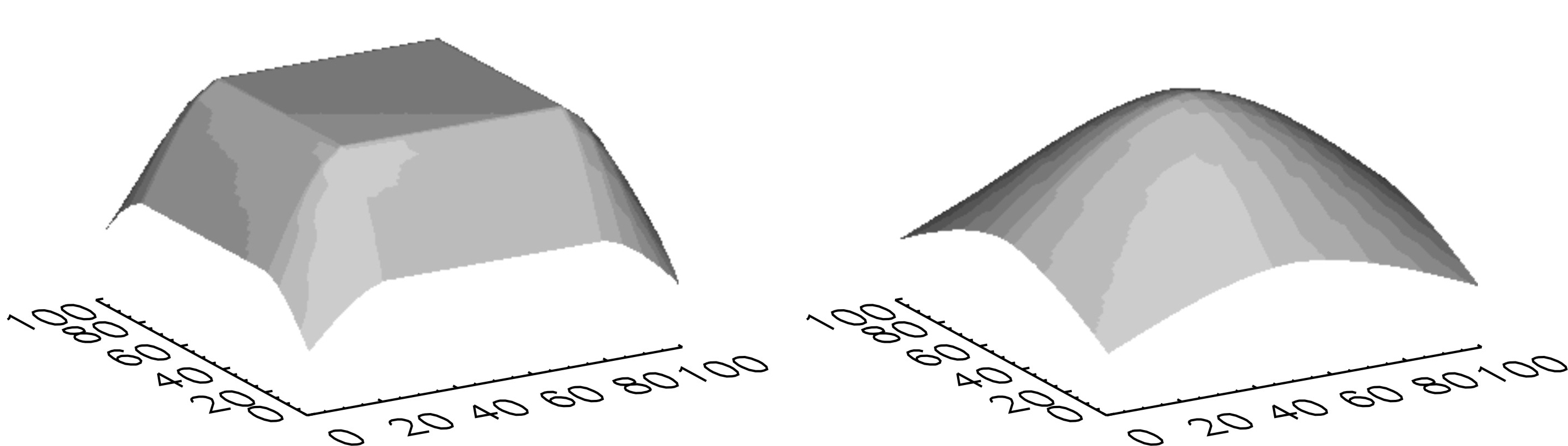}
\end{center}
\caption{Number of re-accreted ejecta on a target modeled by a 100x100 fields chess board like  re-accretion toy model (arb. units). Left: 4 re-accretion zones are added which are 10,20,30, and 40 fields in edge length. Right: re-accretion zones added are between 10 and 100 fields in increments of 10 fields.}
\label{toymodel}
\end{figure}

To support the inhomogeneous target growth of re-accreted particles we use the following toy model. We assume a chess board like matrix as target. We assume each field to be the source of ejecta. We further assume ejecta are re-accreted within a square zone (sub-chess board) centered about the source of ejecta which we call re-accretion zone. Ejecta are taken to be re-accreted with the same probability on each field within the re-accretion zone. According to this scheme we run a loop for all fields of the target and add one particle to each field within the re-accretion zone of a given source field. Fig. \ref{toymodel} shows the result of two such simulations. The target consists of 100 times 100 fields. We add a small number of re-accretion zones in each simulation, which vary in edge length between 10 and 40 fields in the left of Fig. \ref{toymodel} and 10 and 100 fields in the right image, in both cases in increments of ten fields.

The pronounced growth of the target away from the edge is well visualised.  If re-accretion zones are much smaller than the considered target size, then the center part of the target is a plateau (Fig. \ref{toymodel} left). Large re-accretion zones will add curvature to the center (Fig. \ref{toymodel} right). The detailed profile depends on the assumptions of re-accretion probabilities with distance to the impact point or how re-accretion zones relate to ejecta velocities and ejecta angles. The general behaviour of this simple model is consistent with our observations, i.e. cm-size targets start growing similar to Fig. \ref{toymodel} left but also show a significant curvature in the center, though less pronounced than in Fig. \ref{toymodel} right. This is also consistent with the measured ejecta as most ejecta velocities are small and ejecta only travel on mm-scales (Fig. \ref{fragspeed} and \ref{impact}) while larger ejecta velocities exist but with decreasing probability. 
 
Within this simple toy model the target also grows right at the edge, i.e. growth at a vertical wall occurs. The measured targets show no growth at the edge. This is due to erosion which starts from the edges until impact parameters of $70^{\circ}$ are reached. As the inclinations change so will the ejecta angles and re-accretion efficiencies, especially as erosion and less direct sticking get more important everywhere on the target. Independently of the exact growth mechanism the final shape will always be determined by the maximum impact parameter where growth and erosion balance.

It is important to note that both components, direct sticking parts and re-accreted ejecta, occur in individual collisions. They are weighed differently though in their contributions to the total mass added to a given target. For small size targets, e.g. the mm-size cylinder or plate only very slow ejecta can return. On the other side the bottom as extreme target collects all ejecta. Therefore, the topmost surface layer of a grown target will be constituted by the contribution of slow ejecta and fast projectiles in different proportions. Nevertheless, as outlined in the following section, any surface morphology is lost during the evolution of the larger target aggregate. 

\textbf{Porosities / Filling factors:} The dust aggregates grown on all targets measurable had low porosities of 69\% or filling factors of 31\% $\pm$ 3\% on average. This value has been determined as mean filling factor from all targets measured and its standard deviation. Within the error bars this result is independent of the target size, shape or height placed within the chamber. Small targets collecting little ejecta have the same porosity as the chamber bottom collecting all possible ejecta.

With 31\% filling factor the dust aggregates are more compact than aggregates compacted by uni-directional static compression reported by Blum et al. (2006). In their work the filling factor is reaching a plateau of only 20\% filling factor for pressure larger than $10^5$ Pa and for uni-directional compression for the same material used in our experiments. The aggregates reported here are close to the compact aggregates prepared manually for our earlier experiments (Teiser \& Wurm, 2009; Wurm et. al. 2005b). Such aggregates are still more porous than particles in random close packing. So in principle they can be more compact. However, Teiser \& Wurm (2009) and Wurm et al. (2005b) noted in earlier works that large aggregates cannot be compressed further if pressure is not applied from all directions but only on a limited surface area of the dust by a local pressure. A filling factor of $\sim$33\% seems to be the maximum which can be reached for the given kind of dust sample by applying local pressure on a large dust aggregate. Qualitatively, there is a wide plateau in pressure at one maximum filling factor until eventually the target breaks at the boundary where the local pressure is applied. 

As such the maximum filling factor is a unique parameter of a locally compressed dust sample. The value is independent of the history of how the target built up as dust grains are rearranged by the compression. Since the targets grown in our experiments by impacts have filling factors close to this maximum filling factor, they lost all memory of their collisional history. Obviously, dust on the surface gets compressed by the succeeding impacts. It is not important what size-distribution of projectile fragments sticks directly, or what fraction of re-accreted ejecta is placed on the surface. It is not important which grains of an existing surface are removed by an individual collision. The following collisions erase this information, eventually. 

It has to be noted that a highly porous core within a decimetre body might persist, even if the surface layers are compressed by impacting particles. Experimental studies on collisions between compact dust aggregates showed that even in large velocity collisions ($\sim 50$ m/s) only the surface layers are affected (Wurm et al. 2005b, Teiser \& Wurm 2009). Therefore, we argue that, even if a highly porous core exists, the collision dynamics are dominated by the high degree of compression at the surface. The existance of a highly porous core reduces the m/A ratio and therefore also changes the gas coupling time of the body. This will change the relative velocities between particles within the accretion disc, but growth processes as described here will not be affected.

It might be noted that the dust grown on the glass plate edge (see table 1) has a somewhat low filling factor of \mbox{26.4\% $\pm$ 4\%}. This might be due to the small width of the target. For projectiles which are comparable in size to the width of the dust column the impact might no longer be regarded as a local compression. If target dust can move to the sides, this process is comparable to uni-directional compression which results in lower filling factors. As we apply our results to decimetre growth, local compression is the dominating process and low filling factors will not occur.   

\textbf{Critical impact angles:} With statistically more dust growing in the center of a target the surface gets more and more inclined toward the edge and provides different impact angles for further collisions. In our experiments the inclination reaches maximum angles of 70$^{\circ}$ with respect to the normal to the target surface (fig. 6). Therefore, collisional growth occurs for angles at least up to 70$^{\circ}$. If we assume a spherical body in a protoplanetary disc, impact angles larger than 70$^{\circ}$ only occur at a fraction of 12 \% of the total surface cross section hit by projectiles. Steeper surfaces are eroded (see next section), but a randomly rotating body grows on average due to the much larger cross section for sticking collisions. We proved this by using a slowly rotating cylinder as target. Growth occurs at the mantle face of the cylinder here in contrast to the static cylinder front faces (fig.6). A 1.1 mm thick layer grown on the rotating cylinder is shown in fig. 8.

\begin{figure}
\begin{center}
\includegraphics[width=7cm]{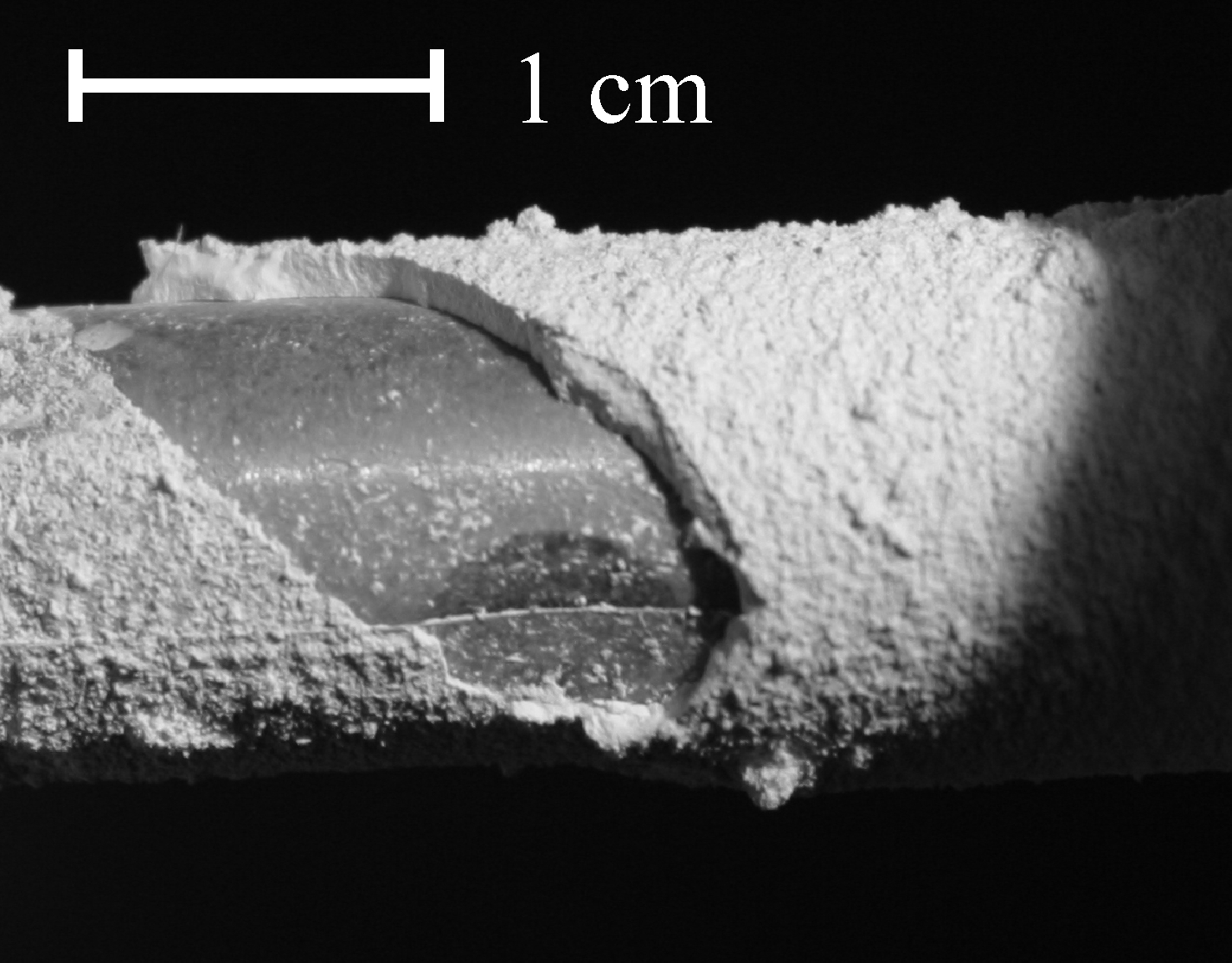}
\end{center}
\caption{Dust crust grown on the mantle face of a cylinder rotating with about 18 rounds per minute.}
\label{rolle}
\end{figure}

The rotating target provides different conditions for sculpturing the topmost layer in contrast to the other targets. Particles which only return by gravity but do not firmly stick to the target yet are forced to leave again by gravity during rotation. Yet, net growth occurs and the same porosity evolves as for the other targets (table 1).

\textbf{Vertical growth:} Fig. 9 shows the dusty surface of a vertical wall of an aluminium block placed in the experiment chamber and a schematic sketch showing the setup of this experiment. The centre of the target is shielded from direct impacts from the top (see schematic sketch). Due to shielding and vertical mounting any growth is due to direct sticking in secondary impacts of dust ejected at the bottom. These aggregates are typically much slower than \mbox{1 m/s} (fig. 5). Closer to the bottom the layer of dust grown is somewhat thicker than further up. This is in agreement to ejecta on average hitting the target more closely to the bottom than further up. At areas which were not protected against direct projectile impacts at \mbox{7.7 m/s,} growth also occurred but erosional streaks are clearly visible in agreement to the observation that growth is possible up to 70$^{\circ}$ impact angle while the vertical walls represent 90$^{\circ}$ impact angle.

\begin{figure}
\begin{center}
\includegraphics[height=5cm]{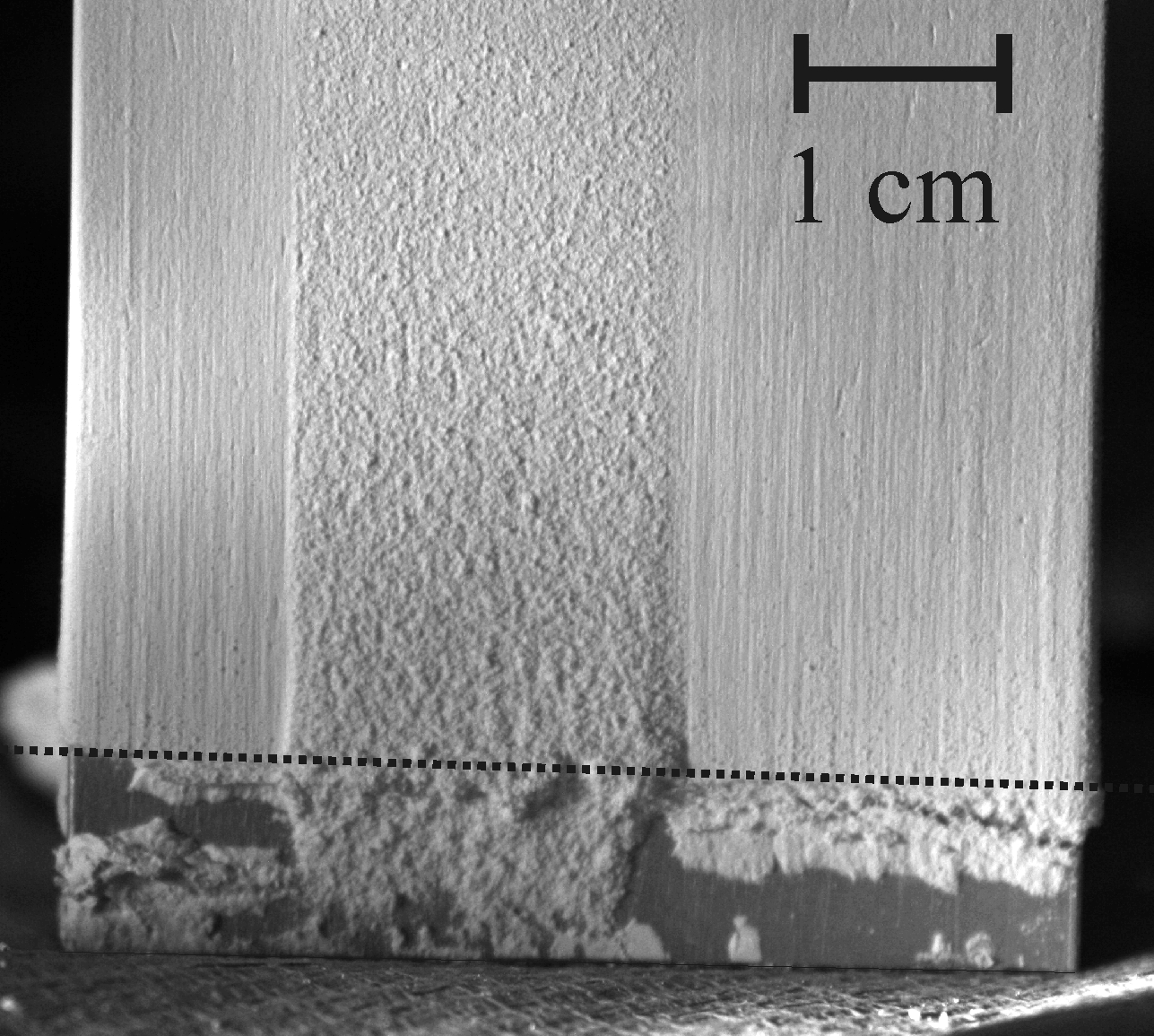}
\includegraphics[height=6cm]{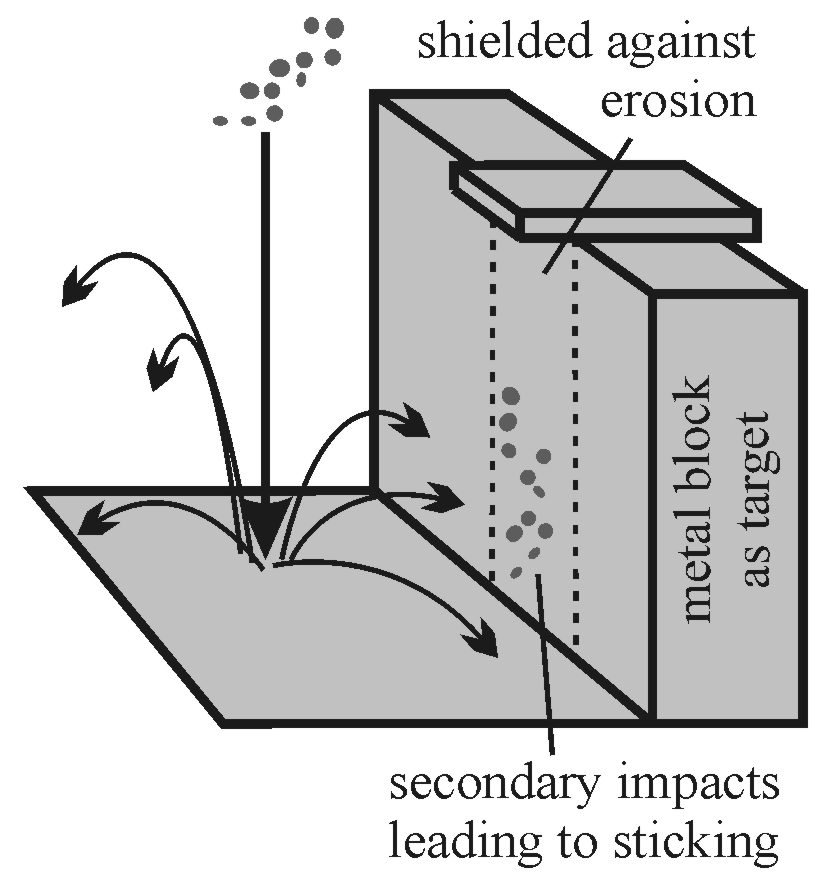}
\end{center}
\caption{Growth on a vertical wall of an aluminium block by accretion of impacting ejecta (result and schematic sketch). In the part shielded against direct impacts agglomerates stick to the wall and form a rough surface. Erosion streaks in exposed parts of the surface are visible.}
\label{seite}
\end{figure}

As the focus of the work presented here is on the growth of aggregates by direct impact of somewhat faster projectiles, a detailed analysis of vertical target growth is beyond the scope of this paper. However, it seems worthwhile to mention that net growth by direct sticking at velocities below or about 1 m/s is possible. Growth might be biased to the low or high end of ejecta velocities or to special ejecta morphologies but this has to be verified in dedicated studies. 

We note that growth below 1 m/s has been observed for fractal dust aggregates (Wurm and Blum 1998) and impacts into highly porous dust aggregates (Langkowski et al. 2008). In individual collisions between arbitrarily picked dust aggregates of similar size only bouncing was observed so far (Blum and M\"unch 1993). In a statistical distribution of randomly generated fragments colliding with slow velocities there is a significant amount of particles which can stick directly. Obviously there are enough fragments sticking which are not removed again by other slow collisions. This is of significance for the formation of dust aggregates of several centimetre, whose further evolution to several decimetre is simulated here.

\section{Gravity and gas drag}

There is little doubt that in the experiments reported here gravity plays an important role in re-accretion of ejected particles. Due to the universal outcome of the experiments it seems not likely though that it is important if an impacting projectile compacts a re-accreted ejecta particle or an aggregate which stuck directly upon impact. It is likely that a typical decimetre dusty body will always evolve in the way observed. Certainly, the detailed accretion efficiencies, which can be divided in the direct sticking part and the re-accreted part, are important for the timescales of growth. We discuss the re-accretion here. While in contrast to our laboratory experiments a small body in a protoplanetary disc has an insignificant self-gravity to attract ejecta, re-accretion occurs by gas drag in a gas flow. This concept has been introduced by Wurm et al. (2001 a,b). There are some similarities and differences between gravity and gas drag with consequences for the evolution of solids in protoplanetary discs. We compare the following two situations of a target and ejecta after a collision here.
\begin{itemize}
\item Gravity: A target is placed in Earth's gravity field and projectiles of smaller size collide with the target top side. Ejecta are accelerated back to the target surface after a collision by Earth's gravity with $g=9.81 \rm m/s^2$. This is the laboratory situation. The experiments reported here show that re-accretion of ejecta occurs at this acceleration even on targets which are only 1 mm in size.
\item Gas drag: The target is placed in a protoplanetary disc. Due to the star's gravity and coupling to the gas it moves relative to the gas of the protoplanetary disc. Viewed from the target it is subject to a head wind. We assume that the target is small enough that the stream lines are directed toward the surface of the target (free molecular flow). Projectiles are smaller than the target and move slower relative to the gas than the target. Projectiles collide with the target from the direction of the head wind (front side). Ejecta are facing the head wind and are accelerated back to the target front surface. Depending on the strength of the gas drag and the target size, ejecta are re-accreted.
\end{itemize}
Gravity and gas drag by the head wind are similar as they can be described by a constant acceleration for a given particle. The acceleration by gas drag is simply given as 
\begin{equation}
a = \frac{v}{\tau}\;.
\label{accelerate}
\end{equation}

Here, $\tau$ is the gas grain coupling time and $v$ is the relative velocity of the ejecta with respect to the gas. The relative velocity $v$ is the sum of the head wind velocity and the ejecta velocity. If the ejecta velocity is small, then $v$ is essentially only the head wind velocity. Therefore, the acceleration is constant to the first order. A constant acceleration results - in analogy to gravity - to ejecta with parabolic trajectories close to the target body.  However, gas drag is selective of ejecta in terms of the gas grain coupling time. In the free molecular flow regime this is given as  (Blum et al. 1996)
\begin{equation}
\tau = 0.68 \frac{m}{A} \frac{1}{\rho_G \nu_G}\;,
\label{tau}
\end{equation}
where $m/A$ is the mass to cross section ratio of the particle, $\rho_G$ is the gas density, $\nu_G$ is the thermal velocity of gas molecules. For a spherical dust aggregate it is
\begin{equation}
\frac{m}{A} = \frac{4}{3} r \rho_P f\;.
\end{equation}
In analogy to our experiments we consider the particle density as $\rho_P=2.6 \rm g/cm^3$. As filling factor $f$ we take the value for compact aggregates of $f=0.31$, keeping in mind that for small aggregates large variations are possible. To specify the gas grain coupling time further, we need to assume a density profile for the protoplanetary disc. We consider the minimum mass solar nebula given by Hayashi et al. (1985).
\begin{equation}
\rho_G = 1.4 \times 10^{-6} \frac{R}{\rm 1AU}^{-11/4} \rm kg m^{-3}\;.
\end{equation}
The temperature is given as
\begin{equation}
T = 280 \left(\frac{R}{\rm 1AU}\right)^{-1/2} \rm K\;.
\end{equation}
The thermal velocity of the gas molecules is given as
\begin{equation}
\nu_G = \sqrt{\frac{8R_{gas}T}{\pi \mu}};
\end{equation}
where $\mu$ is the mean molar mass ($2.34$ g/mol), $R_{gas}$ is the gas constant. Eq. \ref{tau} can then be written as
\begin{equation}
\tau = 3.28 \cdot 10^5 r \left(\frac{R}{\rm 1 AU}\right)^3\;.
\label{tau2}
\end{equation}
With eq. \ref{tau2} and a particle radius of $r= 10 \mu \rm m$ ejecta at \mbox{1 AU} have a coupling time of $\tau = 3s$. To specify the acceleration due to gas drag, we need a head wind velocity (eq. \ref{accelerate}). This also depends on the disc model, i.e. the distance to the star and the target size or more accurate the gas grain coupling time of the target. It is beyond the scope of this paper to give a detailed discussion of all possible re-accretion scenarios of targets of different sizes within a disc. This could not be backed up by the laboratory experiments presented. Here, we concentrate on a decimetre body. In the minimum mass disc discussed the head wind velocities are then of the order of 10 m/s at 1 AU, increasing slowly closer to the star and decreasing at larger distances (Weidenschilling and Cuzzi 1993; Sekiya and Takeda 2003). Given the above values at 1 AU an $ r= 10 \mu \rm m $ particle ejected from a decimetre target is accelerated back to the surface of the target with $a=3.3\, \rm m/s^2$. This is, in fact, comparable to Earth bound gravity conditions and our experiments can immediately be applied here. Such ejecta would return to the target by gas drag only millimetres away from the primary impact side. The distance travelled by an ejected particle before it returns to a surface simply scales with $v_{ej}/a$, where $v_{ej}$ is the ejecta velocity with respect to the target. A decimetre target would still be hit a second time either at a factor 100 less in acceleration or for a factor 100 larger ejecta velocities. As ejecta velocities do not vary that strongly, 10 micron particles would always be re-accreted. 

Other examples of conditions under which re-accretion by a decimetre body is still possible are e.g.  1mm slow ejecta at 1 AU or a 10 micron particle at 4.6 AU. Further out in the disc re-accretion of a decimetre body by gas drag will decrease significantly in efficiency. On the other side larger bodies might re-accrete as the mean free path of the gas increases and larger targets are still in the free molecular regime. Much closer to the star, accelerations get stronger but the mean free path decreases. The mean free path is 6 cm at 1 AU and decreases linearly with density (Sekiya and Takeda 2003). Wurm et al. (2001a,b) showed that re-accretion is still possible if the target is 10 times larger than the mean free path. However, ejecta will be transported around the target instead of back, eventually.

From these considerations it follows that the re-accretion efficiency of ejecta for a decimetre body in the terrestrial planet zone is very high, though detailed analysis is subject to further experiments.  In outer parts of protoplanetary discs the re-accretion efficiencies of decimetre bodies are small. Overall, re-accretion is an important process for dust growth. Its dependence on gas density and ejecta size might induce some biases to growth at various stellar distances, to what end is subject to further research, i.e. global coagulation models.

\section{Discussion}

Experiments by Wurm et al. (2001a,b) showed that growth by gas aided re-accretion of ejecta is possible if small ($<$ 10 $\mu$m) fractal aggregates impact a target of $\sim$100 $\mu$m in size. Scaled to decimetre, these experiments suggested that growth of decimetre bodies should be possible. Here, we use non-fractal dust aggregates of up to about 250 $\mu$m in size as projectiles and larger targets of sizes up to several centimetres. 

With respect to the given velocity and size range it was not clear if the stage from several cm to several dm can really be bridged by collisional growth. Individual collision experiments were not unique in their prediction (Langkowski et al. 2008, Wurm et al. 2005a,b, Blum \& M\"unch 1993). An individual projectile can add mass, just bounce off or remove mass from a target. Here we show that growth of decimetre bodies is indeed the net outcome of repeated random collisions. This assumes that 7.7 m/s is a typical impact velocity and $100 \rm \mu m$ are typical projectile sizes. According to Weidenschilling (1997) a decimetre body at 30 AU gains most of its mass by accretion of small ($50 \mu$m) particles impacting with about 10 m/s, which is consistent with our experiments and was the motivation why we chose these projectile sizes. Another signifcant mass gain of decimetre bodies is due to impacts of larger (1 cm) particles (Weidenschilling 1997). However, impacts of centimetre particles in the given velocity range do not change the target surface in a different way. Bouncing is the dominating process in this parameter regime (Wurm et al. 2005b) and no higher compressions than the maximum of 33\% filling factor can occur (Teiser and Wurm 2009). If collisions in a turbulent disc would occur between similar size decimetre bodies, the situation might change but we speculate that if destruction occurs all fragments would be of similar morphology as compact aggregates. Again, no higher compaction is possible by local or unidirectional compression (Teiser and Wurm 2009).

The incorporation of re-accreted ejecta is a major part of the growth process. The secondary impacts occur only at a moderate velocity of up to 1 m/s. Together with the compaction by succeeding collisions dust gets incorporated into the target. It is likely that this process can be scaled up and down to somewhat lower or larger primary collision velocities though this still has to be verified. We assume that the 7.7 m/s we used is appropriate for decimetre-size bodies colliding with smaller dust aggregates. If so, once aggregates of several cm have grown, the further growth of aggregates up to several decimetres occurs. The experiments therefore bridge another order of magnitude in growth. 

Our results are also important for detailed coagulation models. These need to evolve a size-distribution of solids. The evolution is determined by the outcome of collisions. The outcome of collisions in turn depends on the morphology, i.e. porosity of the dust aggregates. A growing number of experiments detail the outcome of collisions for different parameters (Blum \& Wurm 2008). While these can and should be used for the initial growth phase of about cm-size dust aggregates, they might not be easily incorporated in numerical calculations. Every collision changes the morphology of the dust aggregates, i.e. their porosity. However, the growth beyond mm size and beyond highly porous dust aggregates to more compact configurations occurs rapidly (Wurm \& Blum 2000). At least at the midplane of a disc, it is only a short transient phase. After 1000 y or so, larger bodies will be present. While smaller dust aggregates will as well be present later on, their porosity is no longer determined by growth but fragmentation, as the small aggregates will be ejecta from larger bodies. 

As we showed, larger bodies can be described by a single porosity or filling factor. Even in case of a highly porous body, the collision dynamics are determined by the compression of the outer layers. Fragments, being part of the outer shell of such a body, are likely sharing this porosity as long as sizes are large enough to apply the concept of porosity. Coagulation models can therefore be kept easy for aggregates larger than $\sim$100 $\mu$m by applying a single filling factor of 31\%.  

\section{Conclusions}

Decimetre bodies are of special importance for the evolution of solids in protoplanetary discs. Johansen et al. (2006) recently proposed a scenario of planetesimal formation built on decimetre bodies and gravitational instability. If the major amount of mass could be locked up in decimetre bodies, growth of planetesimals seems feasible under certain assumptions. For detailed modeling the physics of typical compact aggregates (31\% filling factor) might be considered.

In a pure growth model planetesimals form by collisions only. Decimetre bodies are important here as well, as they are the progenitors to metre bodies. The formation of metre-size dust aggregates is considered to be challenging. Collision velocities reach 60 m/s already in laminar or slightly turbulent discs  (Weidenschilling \& Cuzzi 1993, Sekiya \& Takeda 2003). The structure of decimetre bodies as collision targets is of major significance in this context. Teiser \& Wurm (2009) and Wurm et al. (2005b) used compact targets close to the self-consistently generated bodies reported here. They showed that growth at several tens of m/s is possible with such targets even up to 60 m/s (probably more). This allows the formation of planetesimals under certain assumptions. Our experimental results support this model as decimetre bodies are compact. In summary, we find:

   \begin{enumerate}
      \item Bodies easily grow one order of magnitude from several cm to several dm in size at collision velocities of several m/s in collisions with projectiles of 100 $\mu$m size scale. Parts of a projectile stick directly, parts are removed in an individual collision. Ejecta are re-accreted by gas flow under a variety of conditions. 
      \item Independent of the history of the collisions, the decimetre body will have a unique value for the filling factor of 31\% $\pm$ 3\%. 
   \end{enumerate}

The growth efficiency still has to be quantified in detail, depending on the impact parameters (size, impact angle and velocity) and gas flow parameters (velocity, density, target size). The filling factor might also depend in detail on the underlying dust grain size. However, we expect that objects of several decimetres in size grow readily. The collisional history is erased during this growth phase. The first generation of decimetre bodies might still have a core of higher porosity, details being subject to further studies. They might keep a memory of their earlier growth which influences the details of the target dynamics (mass of surface ratio). However, if growth has gone through more generations of destruction and growth of decimetre bodies later on, dust aggregates can be characterised by a single filling factor of about 31\%. This universality offers a solid foundation for further work on the growth of planetesimals. 

\begin{acknowledgements}
      We thank Carsten Dominik for his constructive review of the manuscript. This work was funded by the \emph{Deut\-sche For\-schungs\-ge\-mein\-schaft, DFG\/, FOR 759}.
\end{acknowledgements}

\end{document}